\documentclass[
aps,
pra,
twocolumn,
superscriptaddress,
showpacs,
english,
10pt,
floatfix
]
{revtex4-2}

\usepackage[utf8]{inputenc}
\usepackage{physics}
\usepackage[nohyperlinks,nolist]{acronym}
\usepackage{graphicx}
\usepackage[dvipsnames]{xcolor}
\usepackage{microtype}
\usepackage{bm}
\usepackage{url}
\usepackage{blindtext}
\usepackage{chemformula}
\usepackage[caption=false]{subfig}
\usepackage[version=4]{mhchem}
\usepackage[linktocpage=true,
  colorlinks=true, 
  pdfborder={0 0 0},
  linkcolor=blue,
  citecolor=red,
  filecolor=yellow,
  urlcolor=blue,
  bookmarks,
  pdfauthor={},
]{hyperref}
\usepackage[capitalise]{cleveref}

\newcommand{\UniBasel}{University of Basel, Department of Physics, Klingelbergstrasse 82, CH-4056 Basel, Switzerland}
\newcommand{\uit}{Hylleraas Centre, Department of Chemistry, UiT the Arctic University of Norway, Tromsø N-9037, Norway}
\newcommand{\aalborg}{Aalborg University, Department of Chemistry and Bioscience, Fredrik Bajers Vej 7H, 9220 Aalborg Øst, Denmark}

\newcommand{\mrchem}{\texttt{MRChem}}

\newcommand{\madness}{\texttt{M-A-D-N-E-S-S}}

\newcommand{\half}{\frac{1}{2}}

\renewcommand{\nat}{N_{\text{at}}}
\newcommand{\nel}{N_{\text{el}}}
\newcommand{\etot}{E_{\text{tot}}}
\newcommand{\strain}{\varepsilon_{i j}}
\newcommand{\stress}{\bm \sigma_{ij}}
\newcommand{\kinstress}{\bm \sigma_{ij}^{\text{kin}}}
\newcommand{\xcstress}{\bm \sigma_{ij}^{\text{xc}}}
\newcommand{\elecstress}{\bm \sigma_{ij}^{\text{elec}}}
\newcommand{\decexp}[1]{ \cdot 10^{#1}}
\newcommand{\bde}[2]{$#1 \cdot 10^{#2}$}

\begin{document}

\title{Noise Tolerant Force Calculations in Density Functional Theory: A Surface Integral Approach for Wavelet-Based Methods}

\author{Moritz Gubler} \email{moritz.gubler@unibas.ch} \affiliation{\UniBasel}
\author{Jonas A.\ Finkler} \affiliation{\UniBasel} \affiliation{\aalborg}
\author{Stig Rune Jensen} \affiliation{\uit}
\author{Stefan Goedecker} \affiliation{\UniBasel}
\author{Luca Frediani} \affiliation{\uit}

\begin{abstract}
We introduce a method for computing quantum mechanical forces through surface integrals over the stress tensor within the framework of density functional theory. This approach avoids the inaccuracies of traditional force calculations using the Hellmann-Feynman theorem when applied to multiresolution wavelet representations of orbitals. By integrating the quantum mechanical stress tensor over surfaces that enclose individual nuclei, we achieve highly accurate forces that exhibit superior consistency with the potential energy surface. Extensive benchmarks show that surface integrals over the stress tensor offer a robust and reliable alternative to the direct use of the Hellmann-Feynman theorem for force computations in DFT with discontinuous basis sets, particularly in cases where wavelet-based methods are employed.
In addition, we integrate this approach with machine learning techniques, demonstrating that the forces obtained through surface integrals are sufficiently accurate to be used as training data for machine-learned potentials. This stands in contrast to forces calculated using the Hellmann-Feynman theorem, which do not offer this level of accuracy.
\end{abstract}

\maketitle

\begin{acronym}
\acro{AO}{atomic orbital}
\acro{API}{Application Programmer Interface}
\acro{AUS}{Advanced User Support}
\acro{BEM}{Boundary Element Method}
\acro{BO}{Born-Oppenheimer}  
\acro{CBS}{complete basis set}
\acro{CC}{Coupled Cluster}
\acro{CTCC}{Centre for Theoretical and Computational Chemistry}
\acro{CoE}{Centre of Excellence}
\acro{DC}{dielectric continuum}  
\acro{DFT}{density functional theory}  
\acro{DKH}{Douglas-Kroll-Hess}
\acro{EFP}{effective fragment potential}
\acro{ECP}{effective core potential}
\acro{EU}{European Union}
\acro{FCI}{Full Configuration Interaction}
\acro{GGA}{generalized gradient approximation}
\acro{GPE}{Generalized Poisson Equation}
\acro{GTO}{Gaussian type orbital}
\acro{HF}{Hartree-Fock}  
\acro{HPC}{high-performance computing}
\acro{Hylleraas}[HC]{Hylleraas Centre for Quantum Molecular Sciences}
\acro{IEF}{Integral Equation Formalism}
\acro{IGLO}{individual gauge for localized orbitals}
\acro{KB}{kinetic balance}
\acro{KS}{Kohn-Sham}
\acro{LAO}{London atomic orbital}
\acro{LAPW}{linearized augmented plane wave}
\acro{LDA}{local density approximation}
\acro{MAD}{mean absolute deviation}
\acro{maxAD}{maximum absolute deviation}
\acro{MM}{molecular mechanics}  
\acro{MCSCF}{multiconfiguration self consistent field}
\acro{MPA}{multiphoton absorption}
\acro{MRA}{multiresolution analysis}
\acro{MSDD}{Minnesota Solvent Descriptor Database}
\acro{MW}{multiwavelet}
\acro{NAO}{numerical atomic orbital}
\acro{NeIC}{nordic e-infrastructure collaboration}
\acro{KAIN}{Krylov-accelerated inexact Newton}
\acro{NMR}{nuclear magnetic resonance}
\acro{NP}{nanoparticle}  
\acro{OLED}{organic light emitting diode}
\acro{PAW}{projector augmented wave}
\acro{PBC}{Periodic Boundary Condition}
\acro{PCM}{polarizable continuum model}
\acro{PW}{plane wave}
\acro{QC}{quantum chemistry}  
\acro{QM/MM}{quantum mechanics/molecular mechanics}  
\acro{QM}{quantum mechanics}  
\acro{RCN}{Research Council of Norway}
\acro{RMSD}{root mean square deviation}
\acro{RKB}{restricted kinetic balance}
\acro{SC}{semiconductor}
\acro{SCF}{self-consistent field}
\acro{STSM}{short-term scientific mission}
\acro{SAPT}{symmetry-adapted perturbation theory}
\acro{SERS}{surface-enhanced raman scattering}
\acro{STO}{Slater-Type Orbital}
\acro{WPREL}[WP1]{Work Package 1}
\acro{WPROP}[WP2]{Work Package 2}
\acro{WPAPP}[WP3]{Work Package 3}
\acro{WP}{Work Package}  
\acro{X2C}{exact two-component}
\acro{ZORA}{zero-order relativistic approximation}
\acro{ae}{almost everywhere}
\acro{BVP}{boundary value problem}
\acro{PDE}{partial differential equation}
\acro{RDM}{1-body reduced density matrix}
\acro{SCRF}{self-consistent reaction field}
\acro{IEFPCM}{Integral Equation Formalism \ac{PCM}}
\acro{FMM}{fast multipole method}
\acro{DD}{domain decomposition}
\acro{BLYP}{Becke, Lee, Yang and Parr}
\end{acronym}

\section{Introduction}\label{sec:intro}

\Ac{DFT}~\cite{ksdft1, ksdft2} has become a cornerstone in the computational study of molecular and condensed matter systems due to its balance of accuracy and computational efficiency. Central to the practical application of DFT is the accurate calculation of forces acting on nuclei, which is essential for geometry optimizations~\cite{sqnm,vcsqnm,py_mh}, molecular dynamics simulations~\cite{verlet}, and various other simulations. 
The Hellmann-Feynman theorem~\cite{hellmann_feynman} provides a direct route to force calculations by relating the forces to expectation values of gradients of the Hamiltonian with respect to nuclear positions, thus avoiding the calculation of derivatives of molecular orbitals.

In DFT, the choice of basis sets is crucial for the accuracy and efficiency of calculations. Common basis sets like \acp{GTO}~\cite{gaussorb} offer computational efficiency and ease of integration but cannot fully eliminate basis set errors due to their inability to accurately represent nuclear cusps and their non-systematic nature. 
\Acp{NAO}~\cite{nao} are tuned to resemble physical orbitals and can for example represent nuclear cusps better than \acp{GTO}. The dependence of \acp{GTO} and \acp{NAO} on atomic positions complicates force calculations, rendering the Hellmann-Feynman theorem inapplicable. This necessitates the computation of Pulay forces~\cite{pulay_69}, which is computationally intensive and requires complex corrections. In contrast, systematic basis sets like wavelets and plane waves do not suffer from this issue, making them more efficient for accurate force calculations. Wavelets~\cite{bigdft,madness,mrchem} allow for adaptive resolution and efficient handling of both smooth and sharply varying functions. Plane waves can also be used as basis sets in bulk materials. However, their inability to accurately represent the cusps in the orbitals at the position of the nuclei necessitates the use of pseudopotentials~\cite{pp1,pp2}.

Each of these basis sets has its own advantages and is suited to different types of problems in electronic structure calculations~\cite{elephant}.

% The dependence of \acp{GTO} and \acp{NAO} on atomic positions complicates force calculations, rendering the Hellmann-Feynman theorem inapplicable. This necessitates the computation of Pulay forces~\cite{pulay_69}, which is computationally intensive and requires complex corrections. In contrast, basis sets like wavelets and plane waves do not suffer from this issue, making them more efficient for accurate force calculations.

In this paper, we will focus on \acp{MW}~\cite{Alpert.2002} 
as basis sets which is used in the codes \madness{}~\cite{madness} and \mrchem{}~\cite{mrchem}. Multiwavelets, in the formulation of Alpert~\cite{Alpert.1993} correspond in essence to using polynomials on adaptive grid in such a way that the error is kept rigorously under control\cite{Alpert.2002}. Coupled with the separated representation of the main Green functions\cite{Beylkin.2005t8s} (Poisson and Helmholtz kernels) they enabled \ac{HF}\cite{Yanai.2004k4m} and \ac{DFT}\cite{Harrison.2004df} calculations with this method to achieve very high precision\cite{Jensen.2017}. In twenty years the model has matured from a niche for precise benchmarks on small molecules\cite{Harrison.2004df,Handy.2005} to a robust tool for production calculations for energies\cite{Jensen.2017, Wind.2023, Brakestad.2021} and properties\cite{10.1021/acs.jctc.4c00394,Jensen.2016,Brakestad.2020} including both correlated methods\cite{Valeev.2023, Bischoff.2013} and relativistic Hamiltonians\cite{Brakestad.2024, 10.1021/acs.jctc.3c01056, Anderson.2019ne}. Despite these developments a robust and reliable way to compute gradients (an indispensable tool for any computational chemistry practitioner) has been lacking: the current implementation based on the Hellmann-Feynman theorem requires very high precision to yield acceptable results and is therefore of limited applicability.\cite{fin_nuc} With this development we aim to provide a method to compute geometrical derivative at enhanced precision which will make multiwavelet methods appealing to for quantum chemistry applications.

When multiresolution wavelets are used to represent the Kohn-Sham orbitals, the singularity of the $1/r$ electron-nucleus potential complicates integral evaluations which can be solved by using a smooth form of the nuclear potential~\cite{fin_nuc}. The introduction of this finite, smooth nuclear potential enables the use of the Hellmann-Feynman theorem to compute the forces acting on the nuclei.

However, this approach may encounter noise due to the rapidly changing nature of the approximate nuclear potential and its derivative.
To address these issues, we propose to calculate forces by evaluating surface integrals over the quantum mechanical stress tensor.\cite{force_and_stress,Nielsen_2} This approach mitigates numerical problems inherent in traditional methods through the use of integration surfaces that are far away from all nuclei where the spatial variation of the orbitals is low. It is therefore particularly well-suited for use with wavelet-based DFT methods.
By integrating the stress tensor over a carefully chosen surface that encloses each nucleus, forces can be computed from regions far from the nuclear cusp in the orbitals, thereby avoiding potential numerical issues caused by these cusps.
We believe that stochastic methods like self averaging Kohn-Sham \ac{DFT}~\cite{stoch_ks}  and techniques using discontinuous Galerkin orbitals~\cite{galerkin1,galerkin2} could also benefit from this approach due to its ability to avoid the nuclear cusps.

In this paper, we detail the theoretical foundations of our method, starting with a review of the Hellmann-Feynman theorem and its limitations. We then introduce the concept of the stress tensor in the context of DFT and describe how surface integrals over this tensor can be used to compute forces. Our method is then benchmarked against the traditional approach using a variety of molecular systems, demonstrating its superior accuracy.

\section{Theoretical background}\label{sec:background}
\subsection{Hellmann-Feynman theorem}

The total energy the Kohn Sham density functional theory~\cite{ksdft1,ksdft2} has the following form:
\begin{align}
    \etot = & \sum_{k=1}^{\nel} \bra{\phi_k} \frac{\vb p^2_k}{2 m_e} \ket{\phi_k} - \sum_{k=1}^{\nat} \int \rho(\vb r) \frac{Z_k}{\norm{\vb r - \vb R_k}} d^3 \vb r \nonumber \\
    & + \half \int \int \frac{\rho(\vb r) \rho(\vb r')}{\norm{\vb r - \vb r'}} d^3\vb r d^3\vb r' \nonumber \\
    & + \int \rho(\vb r) \varepsilon_{xc}(\rho, \nabla\rho) d^3\vb r \label{eq:energy}.
\end{align}

The single particle wave functions $\ket{\phi_k}$ are eigenstates of the Kohn Sham Hamiltonian $H_{KS} = \frac{\vb p^2}{2m_e} - \sum_{k=1}^{\nat} \frac{Z_k}{\norm{\vb r - \vb R_k}} + \int \frac{\rho(\vb r'}{\norm{\vb r - \vb r'}} d^3 \vb r' + v_{xc}[\rho] $ that fullfill a single particle Schrödinger equation $H_{KS} \phi_k = \varepsilon_k \phi_k$.
Forces acting on the nucleus $k$ are the negative derivative of \cref{eq:energy} with respect to the position of the nucleus $R_k$.

Derivatives of quantum mechanical expectation values can be calculated using the Hellmann-Feynman theorem\cite{hellmann_feynman} which states that $ \pdv{}{\lambda} \bra{\Psi(\lambda)} H \ket{\Psi(\lambda)} = \bra{\Psi(\lambda)} \pdv{H}{\lambda} \ket{\Psi(\lambda)}$. The Hellman-Feynmann theorem can be used to calculate forces acting on the nuclei within the DFT approximation:

\begin{align}
    \vb F_k &= -\pdv{\etot}{\vb R_k} \nonumber \\
           &= {Z_k \vb E_e(\vb R_k)} - \sum_{i\neq k} Z_i Z_k \left( \frac{\vb R_i - \vb R_k}{\norm*{\vb R_i - \vb R_k}^3} \right)
            \label{eq:hellmann-feynman}.
\end{align}
where $\vb E_e$ is the electronic electric field and $Z_k$ is the proton number of atom $k$.

\subsection{Finite Nucleus Model}

When multiwavelets are used to solve the DFT equations numerically, evaluating the electron-nucleus interaction integrals becomes challenging due to the singularity in the $1/r$ potential. In that case it is more efficient to use an approximate, smooth nuclear potential.

\citet{fin_nuc} propose the use of a smoothed nuclear potential of the form 
\begin{equation}
    u(x) = \frac{\erf (x)}{x} + \frac{1}{3\sqrt{\pi}} \left( e^{-x^2} + 16 e^{-4x^2} \right).
\end{equation}
Here, $ U(r) = \frac{u\left(\frac{r}{c}\right)}{c} $ is the smoothed nuclear potential with the property $\lim_{c\rightarrow 0} U(r) = \frac{1}{r}$. \citet{fin_nuc} also provide a method to estimate the length scale $ c$ of the smoothed potential $U$ to ensure overall precision in the calculations.

The introduction of the approximate nuclear potential $U$ changes the first term in \cref{eq:hellmann-feynman}:
\begin{align}
    \vb F_k =& Z_k \int \int \frac{\vb R_k - \vb r}{\norm{ \vb r - R_k }} U'(\norm{ \vb r - \vb R_k }) \rho(\vb r') d^3 \vb r d^3 \vb r' \nonumber  \\
    &- \sum_{i\neq k} Z_i Z_k \left( \frac{\vb R_i - \vb R_k}{\norm*{\vb R_i - \vb R_k}^3} \right)
            \label{eq:smoot_hellmann-feynman} % this may not be entirely correct check sign
\end{align}
The second term technically also changes, but the difference between $U(r)$ and $\frac{1}{r}$ is insignificant when $r$ is of the order of a bond length.

\subsection{Stress tensor}\label{sec:stress_tensor}

The stress tensor $\stress(\vb r)$ is a symmetric tensor field whose divergence gives the force density:
\begin{equation}
    \label{eq:stress}
    f_i(\vb r) = \partial_j \stress( \vb r).
\end{equation}
The Einstein summation convention is used in \cref{eq:stress} and will be applied throughout the rest of the paper.
A useful property of the stress tensor is that it can be used to calculate the force $\vb F$ acting on a body located in the region of space $\vb V$ by integrating over the surface $S(\vb V)$ of $\vb V$:
\begin{equation} \label{eq:stress_int}
    \vb F_i = \int_{\vb V} \vb f_i(\vb r) d^3 \vb r = \int_{\vb V} \partial_j \stress( \vb r) = \int_{S(\vb V)} \stress n_j da.
\end{equation}
Here, $n_j$ is the j-th component of the normal vector of $S(\vb V)$. In the last step the divergence theorem was used.

The total stress $\langle \stress \rangle$ is an important observable in materials science. It can be used in bulk simulations to determine the optimal lattice shape or drive variable cell shape molecular dynamics simulations. The total stress represents the reaction of a system to a strain deformation $r'_i = (\delta_{ij} + \strain) r_j$, where $\strain$ is the symmetric strain tensor. The total stress is $\left. \pdv{E(\vb R ')}{\strain}\right|_{ \varepsilon = 0}$~\cite{corso94,fast_4g}, where $\vb R'$ contains the strained atomic positions and $E$ is the potential energy of the system.

\begin{figure*}[t]
\centering  
    \subfloat{
        \includegraphics{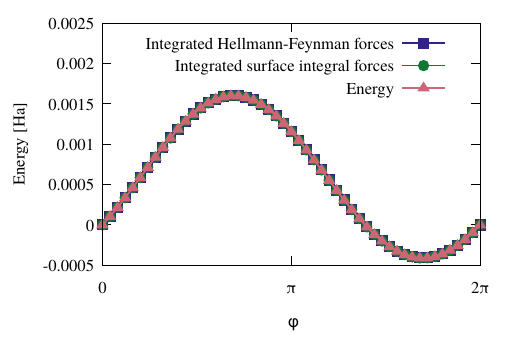}
    }
    \subfloat{
        \includegraphics{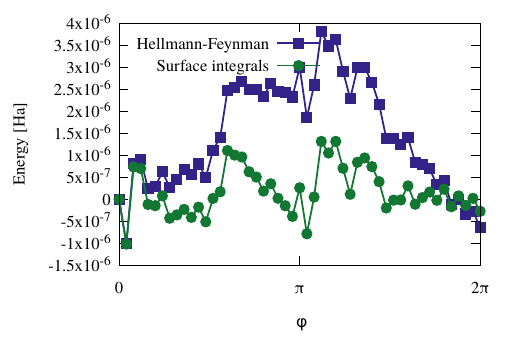}
    }
    \caption{
    Line integration test to check the accuracy of the forces calculated with the Hellmann-Feynman theorem and with the surface integration method presented in this manuscript for a methanol molecule. Details of the test setup are explained in \cref{sec:acc}. On the left, we plot energies along a high dimensional circle, computed directly (red triangles) and as a result of the integration $ E(\varphi = 0 ) - \int_0^{\varphi} \nabla E(\varphi')\cdot \hat{ \vb t}(\varphi') d\varphi'$ for the Hellmann-Feynman (blue squares) and for the stress tensor (green circles) method.
    In the right figure, the difference between the energy along the circle and the corresponding line integral is shown. The smaller difference obtained with the stress tensor method (green circles) indicates that the forces are more consistent with the calculated DFT energy than for the Hellmann-Feynman method (blue squares). Energies were calculated using \mrchem{} with a global precision of $10^{-5}$ and the local density approximation.
    }
    \label{fig:path_int}
\end{figure*}

\section{Method}
In the Born-Oppenheimer approximation, the positions of the nuclei are fixed while the electrons occupy the lowest energy ground state. Due to the variational principle, there is no net force acting on the electrons at any point $\vb r$. By choosing surfaces that enclose only a single nucleus and integrating the quantum mechanical stress tensor over these surfaces, it is therefore possible to compute the forces acting on each nucleus separately, 
%Since there is no net force on the electrons, these surface integrals correspond to the forces acting on the nuclei which can be computed 
using \cref{eq:stress_int}~\cite{force_and_stress}.

The quantum mechanical stress tensor density is the sum of the kinetic stress density $\kinstress$, the Maxwell stress density $\elecstress$ and the exchange correlation stress density $\xcstress$~\cite{force_and_stress,Nielsen_1,Nielsen_2,gauge_stress}.
\begin{align}
    \kinstress(\vb r) =& \frac{1}{4 m_{e}} \sum_{k=1}^{\nel} \left( \pdv{\psi_k^*}{r_i}\pdv{\psi_k}{r_j} + \pdv{\psi_k^*}{r_j} \pdv{\psi_k}{r_i}\right. \nonumber\\
    &- \left. \pdv{\psi_k^*}{r_i}{r_j}\psi_k - \psi_k^* \pdv{\psi_k}{r_i}{r_j} \right) \label{eq:kinstrss} \\
    \elecstress(\vb rr) =& \frac{1}{4 \pi} \left(\vb E_i \vb E_j - \half \delta_{ij} \vb E^2 \right) \label{eq:maxwellstress}\\
    \xcstress(\vb r) = & \delta_{ij} (\epsilon_{xc} - v_{xc}) \rho  - \pdv{(\rho \epsilon_{xc})}{(\pdv{}{r_i}\rho)} \pdv{\rho}{r_j} \label{eq:xcstress}
\end{align}
Here, $\vb E$ is the sum of the electric field from the electrons $\int \frac{\rho(\vb (\vb r - \vb r')}{\norm{\vb r - \vb r'}^3} (\vb r - \vb r') d^3 \vb r'$ and the electric field from the nuclei $\sum_{k=1}^{\nat} Z_k \frac{(\vb R_k - \vb r)}{\norm{\vb R_k - r}^3}$. The second term in \cref{eq:xcstress} accounts for the dependence of $\epsilon_{xc}(\rho, \nabla \rho)$ on the gradient of the charge density in the generalized gradient approximation~\cite{corso94}.

If the orbitals in \cref{eq:kinstrss} are real, the equation can be simplified further. Computing the second derivatives of the wave functions is computationally expensive. \cref{eq:kinstrss} can be rewritten using the product rule such that the second derivatives act on the charge density $\rho$
\begin{equation}\label{eq:better_kin_stress}
    \kinstress(\vb r) = \frac{1}{2 m_e} \left[ \sum_{k=1}^{\nel} 2 \pdv{\psi_k}{r_i}\pdv{\psi_k}{r_j} - \half \pdv{\rho}{r_i}{r_j} \right]
\end{equation}
The expectation value of the stress is given by: $- \frac{1}{m_e} \sum_k \bra{\psi_k} p_i p_j \ket{\psi_k}$. Interestingly, the quantum mechanical kinetic stress density is not given by $- \frac{1}{m_e} \sum_k \psi_k p_i p_j \psi_k$ but by \cref{eq:kinstrss} or \cref{eq:better_kin_stress}. A detailed derivation of the quantum mechanical kinetic stress density is presented by \citet{gauge_stress}.

To evaluate the integral in \cref{eq:stress_int} numerically, one has to define an appropriate surface which contains a single atom. The stress tensor is integrated numerically, and it is therefore  desirable that the integration domain be as smooth as possible, such that the integrand is only varying slowly on that surface. A natural choice that fulfills both conditions is an atom-centered sphere with a radius of half the distance to the nearest neighbour of a given nucleus. This also minimizes possible interferences with the cartesian adaptive grid used for the representation of functions with \acp{MW}.
Another advantage of this choice is that it allows the use of Lebedev integration grids~\cite{Lebedev1977} which give highly accurate results using only a few hundred integration points.
Since the terms needed to compute the stress density are already part of a standard DFT calculation, calculating forces via surface integrals is highly efficient, typically requiring a fraction of the computational time than a single \ac{SCF} iteration. The most expensive computational task is to evaluate the stress density on the integration grid.

\section{Results and discussion}\label{sec:results}

\subsection{Accuracy of the forces}\label{sec:acc}
An important property of forces is their consistency with the corresponding energies. Because the forces are just the negative gradient of the potential energy surface with respect to the atomic positions, that consistency can be checked using line integrals as displayed in \cref{fig:path_int}:
\begin{enumerate}
    \item Generate a circle in the $3\nat$ dimensional configurational space that contains the atomic positions. Let $\varphi\, \in (0, 2 \pi]$ be the polar angle that parametrizes all points on that circle.
    \item Consider the function $E(\varphi)$ which is the potential energy of a point an that circle and $ E(\varphi = 0 ) - \int_0^{\varphi} \nabla E(\varphi')\cdot \hat{ \vb t}(\varphi') d\varphi'$ where $\hat{ \vb t}(\varphi)$ is the unit tangential vector at the point $\phi$. If the forces are the exact gradient of the potential energy surface, those two functions are the same.
    \item Evaluate $E(\varphi)$ on a uniform grid and use these values to solve the integral numerically and compare the two functions in a plot.
\end{enumerate}
In \cref{fig:path_int}, the results of the test are shown for a methane molecule. Energies and forces were calculated with \mrchem{} and a global precision of $10^{-5}$. The difference between the integrated forces and the exact energy is significantly smaller for the new method to calculate forces compared to the previous approach. This demonstrates, that the forces computed with surface integrals over the stress tensor are more accurate compared the previous method. That test was repeated for multiple molecules with the same result.

\begin{figure}[t]
    \centering
    \includegraphics[]{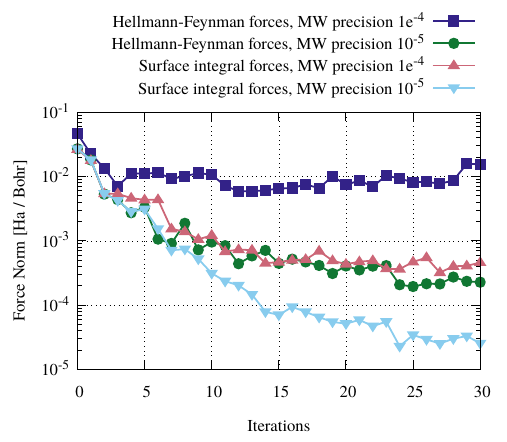}
    \caption{Convergence of the geometry optimizations of an ethanol molecule for the two methods to compute forces using \mrchem{} and the BLYP functional, with different precision thresholds. The stress tensor method is able to achieve a reduction of the residual force norm of an order of magnitude for the sape choice of precision parameter, compared to the Hellmann-Feynman method.
    % The simulations were stopped once all force components were smaller than $10^{-4}$ Hartree per Bohr.
    }
    \label{fig:ethanol_optimization}
\end{figure}

Another useful measure to estimate the accuracy of the forces is presented by \citet{vcsqnm} where one uses the fact the sum of all forces equals zero. However, in a computer simulation the relation is never exactly fulfilled. Let $s_i \ (i=1,2,3)$ be the sum of all $x$, $y$ and $z$ components of the forces respectively. Then, the standard deviation $\sigma$ of the error in the numerically computed forces can be estimated using
\begin{equation}\label{eq:std_dev}
    \sigma = \sqrt{ \frac{1}{3\nat} \sum_{i=1}^3 s_i^2 }.
\end{equation}
This has been done for all the forces that were computed during the benchmark geometry optimizations. The average standard deviation for the forces computed using the Hellmann-Feynman theorem and the forces obtained with surface integrals over the stress tensor is shown in the last row of \cref{tab:res}. The average standard deviation of the force error of forces calculated using surface integrals is almost ten times smaller than forces computed with the Hellmann-Feynman theorem.

\subsection{Geometry optimizations}\label{sec:geopt}
We conducted geometry optimizations for a diverse set of molecules with \mrchem{}~\cite{mrchem} where the new method presented in this article to calculate forces was implemented. In order to test the most general case, it was made sure, that no atoms were on dyadic (coordinates that have the form $m / 2^n$ where $n$ and $m$ are integers) points. All the calculations where done at a global precision of $10^{-5}$ and a geometry optimization was considered converged when all force components were smaller than $10^{-4}$ Hartree per Bohr. The stabilized quasi Newton method~\cite{sqnm, vcsqnm} was used to conduct all geometry optimizations.

Geometry optimizations were conducted with forces using the new method presented in this paper. In all the systems shown in \cref{tab:res}, geometry optimizations using the new forces converged up to a force norm of $10^{-4}$\,Ha per Bohr. This is not the case when the forces are calculated using the Hellman-Feynman theorem. In \cref{tab:res}, the variance of the error in the forces is shown for the 22 test molecules. It is approximately one order of magnitude lower when forces are computed using surface integrals.

\begin{table}[h]
\begin{centering}
\begin{tabular}{l|l|l}
                         & Hellman-              & Surface              \\
                         & Feynman               & Integrals            \\
\hline
\ce{H3Al}                & $1.1 \decexp{-4}$     &  \bde{2.3}{-5}       \\
\ce{H2Be}                & $5.4 \decexp{-6}$     &  \bde{3.0}{-5}       \\
\ce{H3B}                 & $3.5\decexp{-5}$      &  \bde{3.0}{-5}       \\
\ce{C2H6O}               & $1.0 \decexp{-4}$     &  \bde{2.3}{-5}       \\
\ce{H2O}                 & $1.1 \decexp{-4}$     &  \bde{1.2}{-5}       \\
\ce{H2S}                 & $1.6 \decexp{-4}$     &  \bde{3.7}{-5}       \\
\ce{HCl}                 & $9.7 \decexp{-5}$     &  \bde{7.6}{-5}       \\
Hexane                   & $2.9 \decexp{-5}$     &  \bde{9.3}{-6}       \\
\ce{HF}                  & $2.8 \decexp{-5}$     &  \bde{3.2}{-5}       \\
\ce{HLi}                 & $2.8 \decexp{-4}$     &  \bde{1.2}{-6}       \\
Methane                  & $1.9 \decexp{-5}$     &  \bde{1.1}{-5}       \\
\ce{CH4O}                & $6.2 \decexp{-5}$     &  \bde{9.6}{-6}       \\
\ce{H2Mg}                & $9.6 \decexp{-5}$     &  \bde{4.7}{-6}       \\
\ce{HNa}                 & $4.9\decexp{-5}$      &  \bde{4.1}{-5}       \\
\ce{H3N}                 & $4.6 \decexp{-5}$     &  \bde{4.8}{-6}       \\
\ce{H3P}                 & $2.0 \decexp{-4}$     &  \bde{7.4}{-5}       \\
\ce{H4Si}                & $9.5 \decexp{-5}$     &  \bde{8.7}{-6}       \\
Fluoropropylbenzene      & $4.4\decexp{-4}$      &  \bde{4.2}{-5}       \\
Hexanaldehyde            & $2.0 \decexp{-4}$     &  \bde{5.6}{-5}       \\
Hexasilane               & $9.5\decexp{-5}$      &  \bde{2.2}{-5}       \\
Naphtalene               & $1.4 \decexp{-4}$     &  \bde{2.5}{-5}       \\
Nitrobenzene             & $4.0 \decexp{-4}$     &  \bde{2.1}{-5}       \\
                         &                       &                      \\
Average                  & $1.3\cdot 10^{-4}$    &  $2.7\cdot 10^{-5}$  \\
\end{tabular}
\end{centering}
\caption{Force error measure from \cref{eq:std_dev}~\cite{vcsqnm} for 22 test molecules. The last line contains the average force error.}
\label{tab:res}
\end{table}

% In 20 out of the 22 benchmark systems, geometry optimizations converge significantly faster with the newly proposed method to compute forces acting on the nuclei. In the remaining two structures, the convergence rate of both methods is comparable. The average number of iterations is 69 for the forces computed using the Hellmanm-Feynman method~\cite{fin_nuc} and 33 for forces computed using surface integrals over the stress tensor.

In \cref{fig:ethanol_optimization}, the force norm is plotted against the number of geometry optimization iterations. When the forces were calculated using the Hellmann-Feynman method, the force norm stagnates two orders of magnitude before the MW precision. When the newly developed surface integral method is used, the improvement in the force norm is about an order of magnitude with stagnation occurring only one order of magnitude before reaching the MW precision. The error in the forces is a result of the error propagation from \ac{SCF} procedure using \acp{MW}. This shows that the surface integral method leads to a sizeable reduction in error propagation.

\subsection{Comparison with Gaussian orbitals}

\begin{figure}[t]
    \centering
    \includegraphics{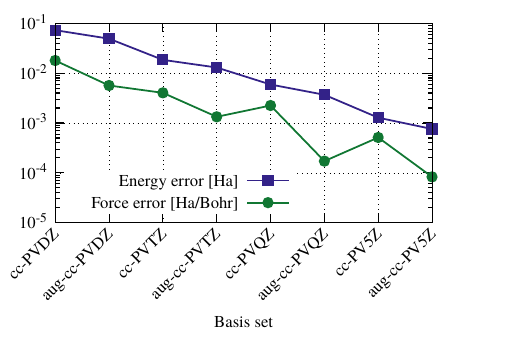}
    \caption{Convergence of Gaussian basis sets compared to \ac{MW} results of a methanol molecule. For the \ac{MW} calculation, a global precision of $10^{-7}$\,Ha was used and the forces were calculated using surface integrals over the stress tensor.}
    \label{fig:gauss_convergence}
\end{figure}

We compared the forces obtained using \acp{MW}, with those obtained using the PySCF~\cite{pyscf} code with the correlation-consistent polarized valence bases (cc-pVnZ)\cite{Nagy.2018}, where n stands for the number of zeta functions used and the presence of the ``aug'' aug indicates that diffuse \acp{GTO} were included in the basis. Methanol was chosen as a test molecule, and the \ac{BLYP}~\cite{b88,lyp} exchange-correlation functional was employed. The methanol molecule was optimized with a MW precision of $10^{-7}$ until all force components were below $10^{-4}$\,Hartree/Bohr. In \cref{fig:gauss_convergence}, for sufficiently large basis sets, the forces from Gaussian orbitals and \acp{MW} were consistent. However, the basis set convergence with Gaussian orbitals was slow and a large number of basis functions had to be used. As expected, augmented bases improve the accuracy of the forces significantly compared to their non-augmented counterpart. For the absolute energy, which is sensitive to the quality of the electronic structure description at the core, the opposite is true: adding valence functions improves the result more than augmentation.

\subsection{Machine learned multiwavelet potential energy surfaces}\label{sec:ml}

\begin{figure*}[t]
\centering  
    \subfloat[Energy correlation\label{subfig:energy}]{
        \includegraphics{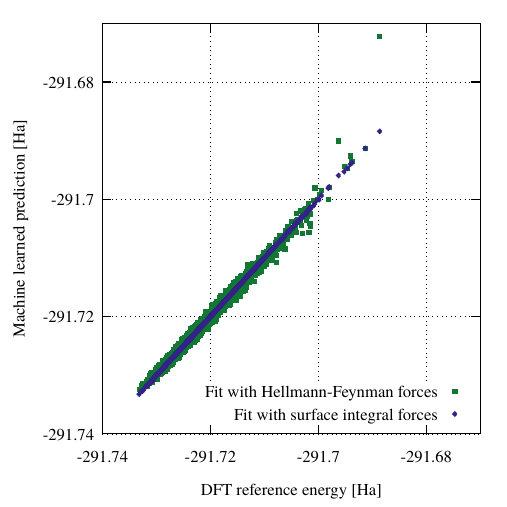}
    }
    \subfloat[Force correlation\label{sufbig:forces}]{
        \includegraphics{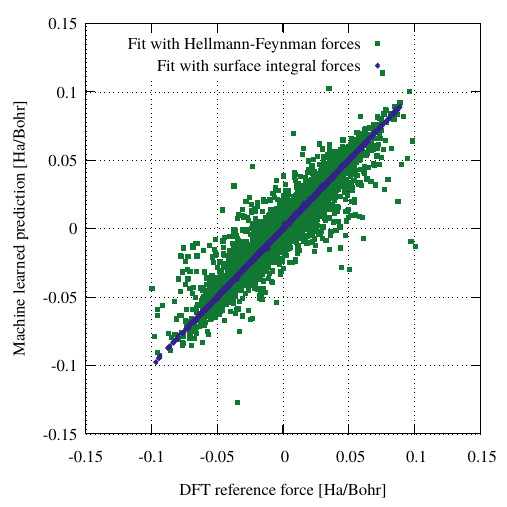}
    }
    \caption{
    Correlation of the machine learned energies (left) and forces (right) against the correct values obtained with DFT for both MLPs.
    }
    \label{fig:ml_correlation}
\end{figure*}

In recent years, machine-learned potentials (MLPs) have seen rapid development, maturing into a robust data driven approach for representing potential energy surfaces. Several MLPs are now commonly used, such as Behler-Parrinello potentials~\cite{2g,4g,fast_4g}, Neural Equivariant Interatomic Potentials (NeqIP)~\cite{nequip}, and Gaussian Approximation Potentials (GAP). These methods have demonstrated their ability to reliably learn both the potential energy surface and its derivatives the forces, making them powerful tools in computational chemistry and materials science.

MLPs typically consist of highly flexible model functions with numerous free parameters designed to predict the quantum mechanical energy of a molecular system. The forces are obtained by taking the negative analytic gradient of this model. To train the model for a specific system, these free parameters are adjusted based on reference data. It is crucial that the forces in this reference data are consistent with the energy since the model assumes this consistency when representing the potential energy surface.

% In order to test if the forces obtained with the the surface integral method presented in this article are precise enough to be used as reference data for MLPs, a machine learned potential using the second generation Behler-Parinello architecture was trained. The silane molecule was chosen as a test system.
To assess whether the forces obtained using the surface integral method are accurate enough for use as reference data in MLPs, a potential using the second-generation Behler-Parinello architecture was trained for the silane molecule.
In order to generate reference data, a molecular dynamics simulation was performed using the FHI-aims~\cite{fhi} code with a time step of 1 femtosecond and a temperature of 500\,K. After 50000 steps, every tenth structure from the last 40,000 steps was added to the dataset, resulting in 4000 structures. These were randomly divided into training (1800 structures), test (200 structures), and validation (2000 structures) sets. The MLP was optimized using the training set, with the test set used to monitor performance and avoid overfitting. The validation set was not used in any way during the training phase and was only used after the training phase to validate the accuracy of the model.

Energies and forces for the training data were calculated using MWs and \mrchem{} with a global wavelet precision of $10^{-4}$\,Ha. Two sets of forces were computed: one using the Hellmann-Feynman method and another using the surface integral method. Correspondingly, two neural network potentials were trained using each set of forces. The training was carried out using the RuNNeR code~\cite{runner1,runner2}.
For both methods to compute forces, a machine learned potential was trained using exactly the same parameters, training, test and validation sets. In \cref{fig:ml_correlation} the correlation between the predicted energies and forces and the corresponding DFT values is illustrated for the validation set.

Comparing with the reference values of the validation set, we find that the accuracy of  the machine learned surface integral forces is significantly better than the accuracy of the Hellmann-Feynman forces. This further demonstrates the superiority  of the method presented in this manuscript. The higher quality of the forces computed with surface integrals also affects the energy correlation as seen in \cref{subfig:energy}. Since the forces are the negative analytic gradient of the machine learned model with respect to the atomic positions, energies can not be trained accurately when the forces are not accurate enough which is the case for the Hellmann-Feynman forces.
For the MLP trained using Hellmann-Feynman forces, the root mean square error (RMSE) of the validation set is $7.7\cdot 10^{-4}$\,Ha for the energies and $5.0\cdot 10^{-3}$\,Ha/Bohr for the forces. In contrast, the MLP trained with surface integral forces achieves an accuracy that is an order of magnitude higher with an RMSE of $4.7\cdot 10^{-5}$\,Ha for the energies and $1.6\cdot 10^{-4}$\,Ha/Bohr for the forces. Notably, the RMSE for the surface integral based MLP approaches the MW precision used in the DFT calculation which was $10^{-4}$\,Ha.

\section{Conclusions}

We have introduced an alternative method for computing forces within the framework of Density Functional Theory, which is particularly suited for scenarios where multiresolution wavelets are employed to represent orbitals and densities. Through extensive benchmarking, we compared our method against the current state-of-the-art technique for force computation. Our results indicate that using the Hellman-Feynman theorem to calculate forces with \acp{MW} is around an order of magnitude less accurate than our newly proposed method that uses surface integrals. % It consistently converges geometry optimizations across all test cases and to serve as reference data for machine learned potentials.

We also performed a line integration test to assess the consistency of the computed forces with the potential energy, and a test to estimate the force error. Our method demonstrated superior accuracy in both tests, significantly outperforming the current approach.

To corroborate the accuracy of \ac{MW} calculations, energies and forces were compared with those obtained using Gaussian basis sets. The results show good agreement when large Gaussian basis sets are employed. However, smaller, commonly used Gaussian basis sets such as cc-pVDZ and cc-pVTZ fail to produce accurate forces in our test. While energy errors might not always impact physical observables due to the fact that only energy differences are physically relevant, forces are directly affected by inaccuracies underscoring the necessity for highly accurate basis sets like \acp{MW} for reliable force calculations.

These findings clearly illustrate that computing forces using surface integrals over the quantum mechanical stress tensor offers substantial advantages over the traditional method based on the Hellmann-Feynman theorem. Our approach not only improves the accuracy of force computations but also enhances the overall efficiency of geometry optimization procedures in \ac{DFT} calculations.

The comparison of machine learned potentials in \cref{sec:ml} demonstrates that, unlike the previous method for computing forces with MWs, the surface integral based approach is accurate enough to serve as training data for machine learned potentials. This enables the creation of machine learned potentials free of basis set errors, using \mrchem{} to generate highly accurate reference data, making it possible to create machine learned potentials with a  previously unattainable level of accuracy.

\subsection*{Code availability}
A reference implementation of the method presented in this paper can be found in the following GitHub repository, and will soon be included in the next \mrchem{} release: \url{https://github.com/moritzgubler/mrchem/tree/surface_forces}

\begin{acknowledgments}
This work is part of a collection of manuscripts for the Festschrift in honor of Prof.~Trygve Helgaker's 70th birthday. I (LF) have had the privilege of spending quite some time with Prof.~Helgaker first as a teacher (European Summer School in Quantum Chemistry (ESQC) in 2001, Sostrup summer school in 2002), and later on as a mentor and collaborator. Prof.~Helgaker has always impressed me for his beautiful lectures and seminars and for his extraordinary capacity to explain challenging science with awesome simplicity. Among the many anecdotes I can recall, I will choose one which fits very well with the topic of our present work. At the ESQC, Prof.~Helgaker lectured about the trust-radius method for geometry optimization: To explain the essence of it, he would challenge a student to find "by hand" the minimum on a potential energy surface (a map on a good old slide for the light board) by covering the map with a piece of paper, which only revealed a circular hole. The student had to discover the minimum by moving the hole around. 

I (LF) would finally like to conclude by thanking Prof.~Helgaker for the many stimulating discussions, and for his contagious passion for the scientific endeavour.

 Financial support was obtained from the Swiss National Science Foundation (project 200021 191994), the Deutsche Forschungsgemeinschaft in the framework of the DFG priority program SPP 2363, and the Research Council of Norway through its Centres of Excellence scheme (Hylleraas centre, 262695). The calculations were performed on the computational resources of the Swiss National Supercomputer (CSCS) under project s1167, on the Scicore (\url{http://scicore.unibas.ch/}) computing center of the University of Basel and on the the Norwegian Metacenter for Computational Science (NOTUR) infrastructure under the nn14654k grant of computer time.
\end{acknowledgments}

\bibliographystyle{apsrev4-2}
\bibliography{main}

\end{document}